\documentclass[a4paper]{article}
\usepackage{authblk}
\usepackage[autostyle=true]{csquotes}
\usepackage{url}
\usepackage{graphicx}
\usepackage{lineno}
\usepackage{bm}
\usepackage{cite}
\usepackage[a4paper]{geometry}
\usepackage{amsmath}
\usepackage{caption}
\usepackage{subcaption}
\usepackage{epstopdf}
\usepackage{placeins}
\usepackage{url}
\usepackage{todonotes}
\usepackage{tikz}
\usetikzlibrary{calc,
                positioning}
\usepackage{tkz-euclide}
\usepackage{mwe}

\newcommand{\R}{\bm{R}}

\newcommand{\E}{\bm{E}}

\newcommand{\HH}{\bm{H}}
\newcommand{\Einc}{\bm{E}^{\rm inc}}

\newcommand{\Etot}{\bm{E}^{\rm tot}}

\newcommand{\x}{\bm{x}}
\newcommand{\y}{\bm{y}}

\DeclareMathOperator\dist{dist}
\graphicspath{Images/}

\title{Rethinking photonic nanojets: a new definition and design paradigm}
\author[1]{Mirza Karamehmedovi\'c  \thanks{Corresponding author, mika@dtu.dk}}
\author[1]{Kristoffer Linder-Steinlein}
\author[2]{Jesper Gl\"uckstad}
\affil[1]{Department of Applied Mathematics and Computer Science, Technical University of Denmark, DK-2800 Kgs. Lyngby, Denmark}
\affil[2]{SDU Centre for Photonics Engineering, University of Southern Denmark, DK-5230 Odense, Denmark}
\date{}                     
\begin{document}
\maketitle

\begin{abstract}
We propose a rigorous, physically interpretable, and quantifiable definition of the photonic nanojet (PNJ). This framework resolves longstanding ambiguities in measuring PNJ dimensions and leverages an optimal mass transport-based metric to quantify PNJ quality. Building on this metric, we develop a PNJ steering methodology that requires no opto-mechanical intervention, relying solely on phase-only illumination modulation.
\end{abstract}

\section{Introduction}

\begin{figure}[hbt!]
\centering
\includegraphics[scale=0.5]{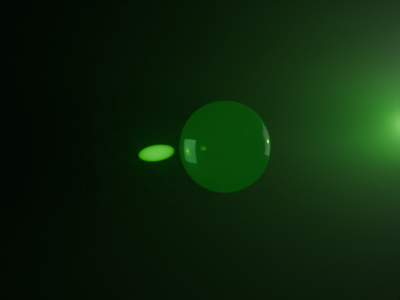}
\caption{Artistic rendering of a photonic nanojet created by a dielectric micro-element.}
\label{fig:art}
\end{figure}

Photonic nanojets (PNJs)~\cite{Darafsheh-2021,2023-phase-only_PNJ,2022-PNJ1,Heifetz-2006,2022-SPIE,Lecler-2019,Itagi-2005} are highly localized, intense light beams that form in the shadow region of a dielectric microstructure when it is illuminated by coherent light. Figure~\ref{fig:art} illustrates the phenomenon.
The high field intensity and narrow width of the PNJ focus, relative to the illuminating wavelength, enable highly localized measurements and applications, including super-resolution optical microscopy~\cite{Huszka-2019,Chen-2004,Li-2005,Darafsheh-2012}, micromanipulation~\cite{Rodrigo,Minin-2020,Neves-2015,Rodrigo2}, as well as fluorescence and Raman microscopy~\cite{Sergeeva-2020,Ruzankina-2020}. Consequently, PNJs remain an intensely studied subject both theoretically and experimentally. We provide a more comprehensive literature overview in~\cite{2023-phase-only_PNJ}.

\begin{figure}
    \centering
    \includegraphics[width=0.45\linewidth]{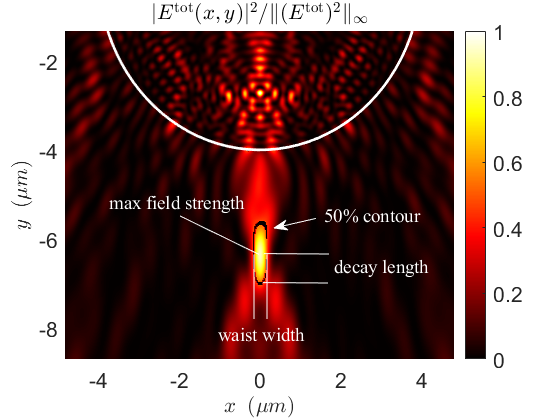}
    \includegraphics[width=0.45\linewidth]{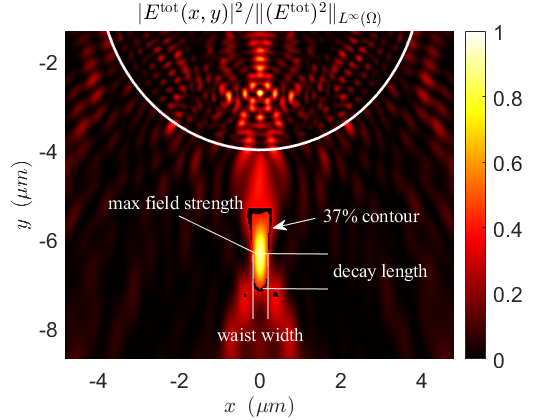}
    \caption{The commonly used PNJ measures of waist width and decay length are intuitive and useful, but the underlying choice of the field intensity level curve is not rigorously justified. Left: a PNJ with FHWM measures. Right: the same PNJ with $e^{-1}$ measures.}
    \label{fig:hand-waving}
\end{figure}

\begin{figure}
    \centering
    \includegraphics[width=0.31\linewidth]{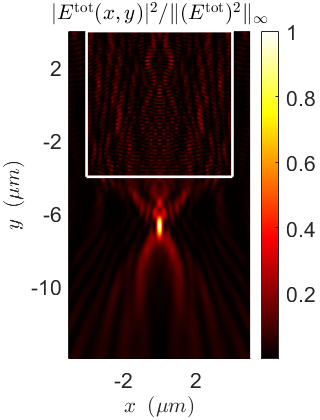}
    \includegraphics[width=0.31\linewidth]{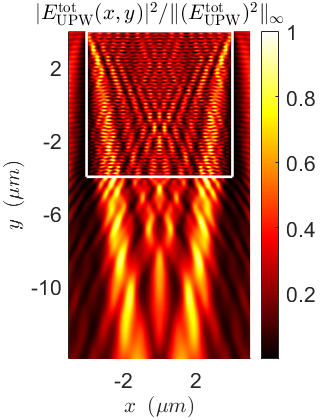}\\
    \includegraphics[width=0.31\linewidth]{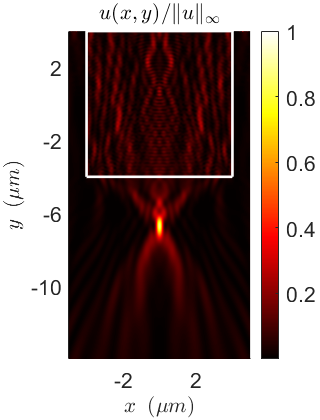}
    \includegraphics[width=0.31\linewidth]{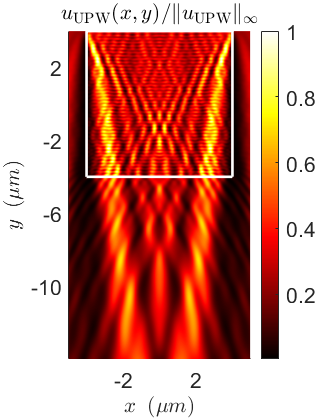}    
    \caption{A highly localized and a delocalized field produced by two different illuminations (computed structured illumination in the left column and a uniform plane wave illumination in the right column) of the same micro-element. The $L^{\infty}$ normalization is performed over the shown domain.}
    \label{fig:energy}
\end{figure}

The investigation of PNJs and their application require a rigorously justified, physically meaningful, and quantifiable definition of what a PNJ actually is. So far, this has lacked in the PNJ community, notably because of the greater problem of rigorously defining and measuring the field concentration phenomena occurring in general in such diverse branches of physics as quantum mechanics, electrodynamics, acoustics, and many more. Specificially, PNJs were so far characterized in terms of the geometry of the associated $e^{-1}$ or full width at half maximum (FWHM) level curves for the electric field intensity (or the square of the electric field intensity), yielding measures such as waist width and decay length, see Figure~\ref{fig:hand-waving}.
While such measures are immediately accessible and physically sensible, they are arbitrary in the sense that no rigorous result requires a particular choice of any field level curve. Moreover, it is difficult to use the waist width and the decay length directly as control parameters in PNJ design. An essential issue determining the applicability of PNJs in, e.g., super-resolution optical microscopy is the ability to control them rapidly and accurately. Recent innovations in this area include the use of optically trapped levitating nanoparticles~\cite{Sergeeva2020A, Hua2023, Simon2024}, shaped optical fiber tips~\cite{Pierron2019,Bouaziz2021,Aljuaid2022,Vairagi2023,Yue2023,Umar2023,Li2024}, and metallic masks~\cite{Liu-2021}. Achieving precise, rapid steering of PNJs holds promising potential for applications such as selective photo-switching of closely spaced sites along biomolecules~\cite{Sergeeva2020B,LocalizedPNJ2022,Xiong2024} and unlocking the full capabilities of label-free, microsphere-assisted optical super-resolution microscopy. In~\cite{2022-PNJ1,2023-phase-only_PNJ} we proposed and demonstrated numerically such PNJ control in 2D using modulated illumination~\cite{GluckstadMadsen-2023} of simple micro-elements, hence with no opto-mechanical intervention. We furthermore estimated theoretical limits to PNJ size in~\cite{2022-PNJ1,2018-bandwidth}. As an illustration of our results in~\cite{2022-PNJ1,2023-phase-only_PNJ}, Figure~\ref{fig:energy} shows two electric field intensities, and the associated time-average electromagnetic field densities, produced by the same micro-element and different illuminations, one field clearly more localized than the other.
However, instead of directly minimizing the PNJ waist width and decay length, it was here necessary to rely on an indirect formulation of a "PNJ field." Finally, PNJ design based on the maximization of the electromagnetic power within a specified region was reported in, e.g.,~\cite{Paganini-2015}, but such approaches also do not directly define nor optimize the field concentration.

In~\cite{2024-localization_improved_concept} we introduced a new, mathematically justified way of measuring the localization of functions, and specifically of solutions of partial differential equations. We here develop this notion further in the direction of applications to time-harmonic electromagnetic fields, then propose a new, rigorously justifed and quantifiable definition of the photonic nanojet, and finally demonstrate its use in the rapid radial and lateral steering of PNJs without opto-mechanical intervention that uses simple homogeneous micro-elements and a machine learning-aided algorithm for the computation of the phase-only modulation of the incident field.

In Section~\ref{sec:new_measure} we introduce and justify rigorously our new measure for photonic nanojets. Then, in Section~\ref{sec:steering}, we use the new measure together with a machine learning-aided optimizer to achieve PNJ steering over a large dynamical range and with no opto-mechanical intervention. Finally, we offer our conclusion in Section~\ref{sec:conclusion}.

\section{A new definition and measure of quality of photonic nanojets}\label{sec:new_measure}

In this section we adapt our definition of localization of probability densities in subdomains of $\R^d$ from~\cite{2024-localization_improved_concept} to the case of concentration of electromagnetic fields and energy densities as observed in PNJs. The resulting new definition of the photonic nanojet is rigorous and quantifiable, and, crucially, it directly helps to control the spatial location of PNJs by providing a relevant objective function to be minimized. Recall that in a lossless, source-free medium with the electric permittivity $\varepsilon$ and the magnetic permeability $\mu$, the time-average total electromagnetic energy density $u(\x)$ associated with a time-harmonic electromagnetic field $(\E,\HH)$ is given by~\cite[pp. 26--27]{Balanis}
\[
u(\x)=\frac{1}{4}\left(\varepsilon|\E(\x)|^2+\mu|\HH(\x)|^2\right)\quad[\rm{Joule}/m^3].
\]
Obviously, $u(\x)\ge0$. The idea now is to use the notion of optimal mass transport~\cite{Villani} to quantify the concentration of $u(\x)$. Specifically, given an electromagnetic field $(\E,\HH)$ in a domain $\Omega\subseteq\R^d$ with Lebesgue measure $|\Omega|=\int_{\Omega}d\x$, we ask for the optimal cost of mass transport from the least localized (flat) unit-mass profile in $\Omega$, the function $g(\x)=1/|\Omega|$, to the profile \[
\widetilde u(\x)=\frac{u(\x)}{\int_{\Omega}u(\y)d\y}
\]
attained by the time-average electromagnetic energy density associated with the field $(\E,\HH)$ and normalized to unit mass in $\Omega$. This optimal cost is given by the Wasserstein-2 distance $W_2(\mu_{\widetilde u},\nu_g)$ between the multiplicative measures $\mu_{\widetilde u}(\x)=\widetilde u(\x)d\x$ and $\nu_g(\x)=g(\x)d\x$ on $\Omega$, defined by~\cite[Definition 6.1, p. 93]{Villani}
\[
W_2(\mu_{\widetilde u},\nu_g)=\sqrt{\inf\left\{\iint_{(\x,\y)\in\Omega}|\x-\y|^2d\gamma(\x,\y),\,\,\gamma\in\Gamma(\mu_{\widetilde u},\nu_g)\right\}}
\]
with $\Gamma(\mu_{\widetilde u},\nu_g)$ the set of couplings between $\mu_{\widetilde u}$ and $\nu_g$,
\[\int_{\y\in\Omega}d\gamma(\x,\y)=\mu_{\widetilde u}(\x),\quad\int_{\x\in\Omega}d\gamma(\x,\y)=\nu_g(\y),
\]
and with the quadratic transport cost function $c(\x,\y)=|\x-\y|^2$. More concentrated energy density profiles require higher transport costs $W_2(\mu_{\widetilde u},\nu_g)$. Given the center of mass of $\widetilde u$,
\[
\x_{{\rm cm}(\widetilde u)}=\frac{1}{|\Omega|}\int_{\Omega}\x\widetilde u(\x) d\x,
\]
the standard deviation 
\[
\sigma(1/|\Omega|)=\sqrt{\frac{1}{|\Omega|}\int_{\Omega}|\x-\x_{{\rm cm}(\widetilde u)}|^2d\x}=\sqrt{\int_{(\x,\y)\in\Omega}|\x-\y|^2\frac{1}{|\Omega|}\delta(\y-\x_{{\rm cm}(\widetilde u)})d\x d\y}
\]
is the optimal mass transport cost of transforming the profile $1/|\Omega|$ to the Dirac delta $\delta(\cdot-\x_{{\rm cm}(\widetilde u)})$ with singularity at $\x_{{\rm cm}(\widetilde u)}$, which is the highest-concentrated unit-mass energy density profile in $\Omega$ with the same center of mass as $\widetilde u$. We now define the relative concentration coefficient of the electromagnetic field $(\E,\HH)$ in $\Omega$ as
\begin{equation} \label{eqn:pnj_meas}
 C(\E,\HH;\Omega)=\frac{W_2(\mu_{\widetilde u},\nu_g)}{\sigma(1/|\Omega|)}\in[0,1),   
\end{equation}
and, given a PNJ-like field in $\Omega$, we quantify its quality by computing $C(\E,\HH;\Omega)$. Similarly, if we wish to design an illumination that produces a PNJ in $\Omega$ then we do this by maximizing $C(\E,\HH;\Omega)$, where we set the desired center of mass (PNJ focus) $\x_{\rm cm}\in\Omega$ ourselves. Our definition of field concentration is consistent with the intuition that uniform plane waves are maximally delocalized non-trivial solutions of time-harmonic Maxwell's equations. Indeed, for such waves both $|\E|$ and $|\HH|$ are constant, leading to $\widetilde u=1/|\Omega|=g$, thus to $W_2(\mu_{\widetilde u},\nu_g)=0$, and finally to $C(\E,\HH;\Omega)=0$.

As we demonstrate and discuss in~\cite{2024-localization_improved_concept}, certain complications can arise in the optimal mass transport interpretation when the 'mass' of the electromagnetic energy density is close to the boundary $\partial\Omega$ of the domain $\Omega$. This is because the cost of transporting that mass to the constant profile $1/|\Omega|$ is higher than that of transporting the mass near the center of $\Omega$, which is free to move in any direction. However, as evidenced by the numerical results of Section~\ref{sec:steering}, we find a way to alleviate this issue using a modification of the original quadratic transport cost function $c(\x,\y)$. Specifically, we define the boundary-compensated cost function
\begin{equation*}
c_{\rm bdy}(\x,\y) = \begin{cases}
                \min \left(\dist(\x,\partial\Omega), \dist(\y,\partial\Omega)\right), & \dist(\x,\partial\Omega) > \tau\,\,\,\text{and}\,\,\, \dist(\y,\partial\Omega) > \tau, \\
                0, & \text{otherwise},
            \end{cases}
\end{equation*}
where $\tau$ is a parameter that controls the thickness of a band around $\partial\Omega$ where the transport cost is set to zero. The corresponding optimal transport costs is now
\[
W_{{\rm bdy}}(\mu_{\widetilde u},\nu_g)=\sqrt{\inf\left\{\iint_{(\x,\y)\in\Omega}c_{\rm bdy}(\x,\y)d\gamma(\x,\y),\,\,\gamma\in\Gamma(\mu_{\widetilde u},\nu_g)\right\}}
\]
and
\[
\sigma_{\rm bdy}(1/|\Omega|)=\sqrt{\frac{1}{|\Omega|}\int_{\Omega}c_{\rm bdy}(\x,\x_{{\rm cm}(\widetilde u)})d\x}.
\]
Our boundary-compensated relative concentration coefficient for the electromagnetic field $(\E,\HH)$ in $\Omega$ is then
\begin{equation} \label{eqn:pnj_meas_comp}
 C_{\rm bdy}(\E,\HH;\Omega)=\frac{W_{\rm bdy}(\mu_{\widetilde u},\nu_g)}{\sigma_{\rm bdy}(1/|\Omega|)}.
\end{equation}
Let us finally mention that, instead of concentrating the time-average electromagnetic energy density, one can optimize the concentration of the modulus of the total electric field or of the total magnetic field. In this work we choose to use the energy density, motivated by a numerical investigation of the relative performances of the approaches.

\section{A machine-learned PNJ steering algorithm}\label{sec:steering}
\begin{figure}[!htbp]
    \centering
    \includegraphics[scale=0.5]{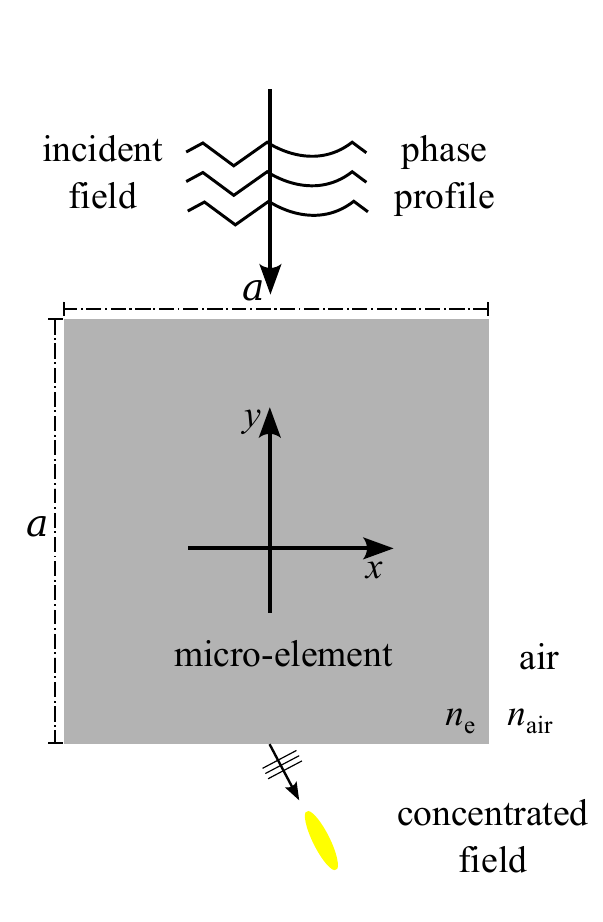}
    \caption{Diagram of the setup. The shaded area is a micro-element with refractive index $n_{\rm e} > n_{\rm air}=1$. The incident field propagates in the negative $y$-direction. The three curvy lines illustrate the phase modulation $\varphi(x)$ of the incident wave. Near the bottom of the micro-element, the PNJ is depicted as the concentrated field.}
    \label{fig:scheme}
\end{figure}

In this section, we use the boundary-compensated PNJ measure \eqref{eqn:pnj_meas_comp} as an objective function in an optimization with the goal of steering a PNJ to a given location relative to a simple microelement. Specifically, we choose a domain $\Omega$ (for example, a small disk or rectangle) where we want a high electromagnetic energy concentration, and select a desired PNJ focus point $\bm{x}_{cm}\in\Omega$, say at the center of $\Omega$. The goal is now to maximize the value $C_{\rm bdy}(\E,\HH;\Omega)$ from~\eqref{eqn:pnj_meas_comp} as a function of a phase-only modulated incident field. We use as the incident field a time-harmonic plane wave propagating in the negative $y$-direction, $\Einc(x,y)=\widehat{\bm{z}}e^{jk_0y}e^{j\varphi(x)}$; see Figure \ref{fig:scheme}, where the three curvy lines illustrate an interpolated version of the phase modulation $\varphi(x)$, defined at the top boundary of the microelement. 
\\This amounts to solving the optimization problem
    \begin{equation}\label{eqn:mini}
        \begin{split}
            & \min_{\varphi}\,\,C_{\rm bdy}(\bm{E}(\varphi),\bm{H}(\varphi); \Omega)^{-1} \quad \text{with the given fixed}\,\,\bm{x}_{cm},
        \end{split}
    \end{equation}
which can be done in many ways. Since the main focus of this article is the definition of PNJs in terms of the relative transport cost \eqref{eqn:pnj_meas_comp}, the method of solution of the optimization problem \eqref{eqn:mini} is less important. We investigated numerically the performance of standard methods such as the Nelder-Mead, implemented in, e.g., Matlab and SciPy, using either \texttt{fminsearch} (Matlab) or the \texttt{minimize} (SciPy) subroutine. We ultimately found that framing the problem in terms of Kolmogorov-Arnold networks (KANs) worked best, and for completeness we now include a brief rendering of the KAN framework. The KAN networks intuitively relate to fully connected multilayer perceptrons in how the input propagates to the output. However, rather than having learnable weights on the edges, the activation functions on the nodes are learnable. We refer to \cite{Liu2024}  for an in-depth discussion of KANs. We assume that the phase function $\varphi$ can be represented as
    \begin{equation*}
        \varphi(\bm x) = \sum_{q=1}^{2n+1} \Phi_q \left( \sum_{p=1}^{n} \phi_{q,p}(x_p)\right),\quad\x=(x_1,\dots,x_p),
    \end{equation*}
for a single-layer KAN. For a KAN with $n_{\rm in} > 1$ and $n_{\rm out} > 1$, each layer can be represented as a matrix,
    \begin{equation*}
        \Phi = \lbrace \phi_{q,p} \rbrace, \quad p = 1,2, \cdots, n_{\rm in}, \quad \text{and} \quad q = 1,2, \cdots, n_{\rm out},
    \end{equation*}
for some functions $\phi_{q,p}: \text{input} \to \R$. For KANs, the functions $\phi_{q,p}$ have trainable parameters and are represented by splines. Finally, for an $L-$layer KAN, the representation will be the composition of each individual $\Phi$. Using this, we can formulate the minimization problem \eqref{eqn:mini} as an unsupervised neural network by now training the KAN to find the parameters of $\phi_{q,p}$ minimizing $C_{\rm bdy}(\bm{E}(\varphi),\bm{H}(\varphi); \Omega)^{-1}$. This approach is only implicitly dependent on dimensionality, and with only minor changes the KAN can be adapted to any dimension.
%
%
%
%
%
%

Regardless of the employed optimization method, we saw a trade-off between the size of the domain $\Omega$ within which we forced the electromagnetic energy to concentrate on the one hand, and the achieved precision in the positioning the PNJ on the other hand. The smaller the volume (Lebesgue measure) of $\Omega$, the more regularized the optimization problem became. However, with $\Omega$ too small, we did not allow enough field variation within the domain, so the effect of variation of the phase of the incident field had little to no effect on what was observed within the computational domain. Another regularization of our optimization problem~\eqref{eqn:pnj_meas_comp} came from the fact that we sought a solution within the set of electromagnetic energy densities produced by phase-only modulated illuminations of a particular microelement geometry and refractive index. Finally, symmetry constraints would further regularize the optimization problem, for instance if the desired PNJ was placed symmetrically with respect to the given microelement.

\begin{figure}[hbt!]
    \centering
    \includegraphics[width=\textwidth]{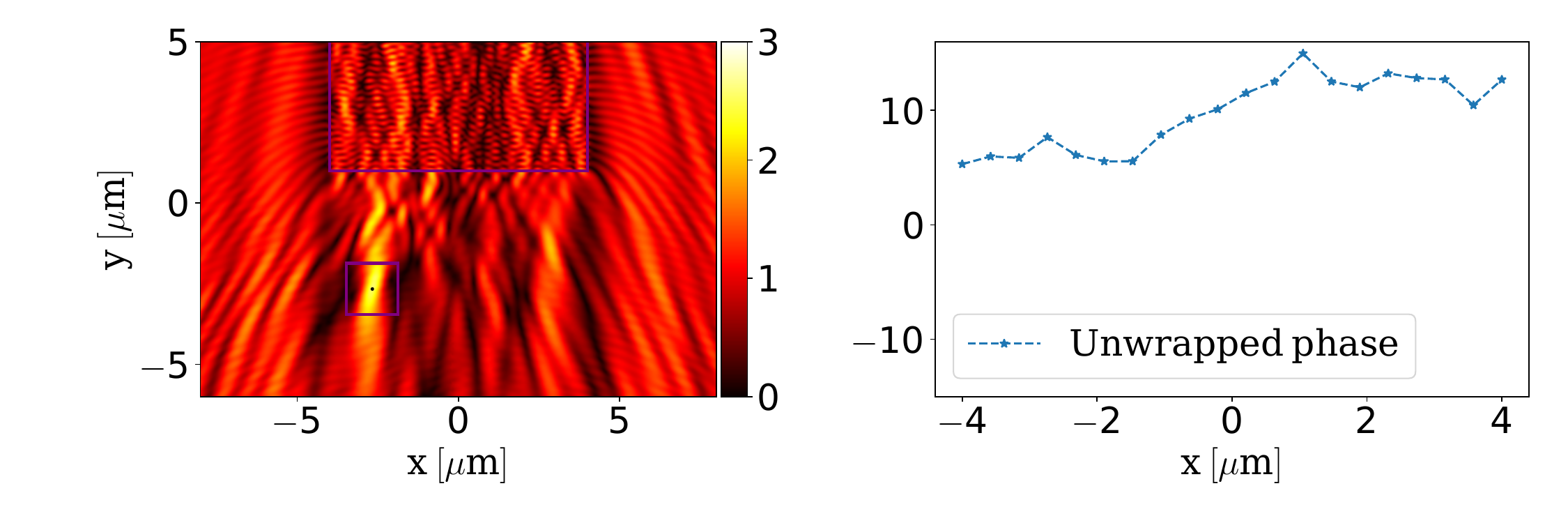}\\
    \caption{An optimized, 20-component phase modulation $\varphi(x)$, and the resulting $|\Etot|$. The PNJ position is 3.67 $\mu m$ along the negative $y$-axis from the micro-element boundary and at $x=-2.67\,\mu$m. The phase is embodied with a resolution given by $\lambda / (2 N_A)$, with $N_A$ the numerical aperture.}
    \label{fig:pos33}
\end{figure}

\begin{figure}[hbt!]
    \centering
    \includegraphics[width=\textwidth]{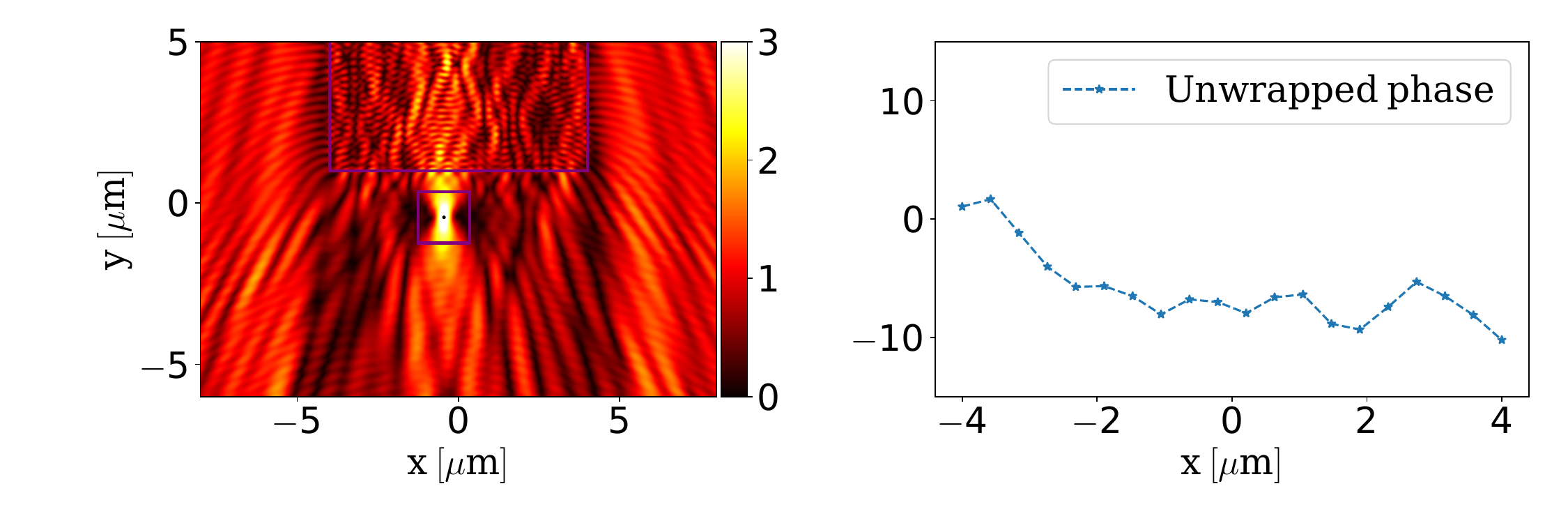}
    \caption{An optimized, 20-component phase modulation $\varphi(x)$, and the resulting $|\Etot|$. The PNJ position is 1.44 $\mu m$ along the negative $y$-axis from the micro-element boundary and at $x=-0.44\,\mu$m. The phase is embodied with a resolution given by $\lambda / (2 N_A)$, with $N_A$ the numerical aperture.}
    \label{fig:pos88}
\end{figure}

\begin{figure}[hbt!]
    \centering
    \includegraphics[width=\textwidth]{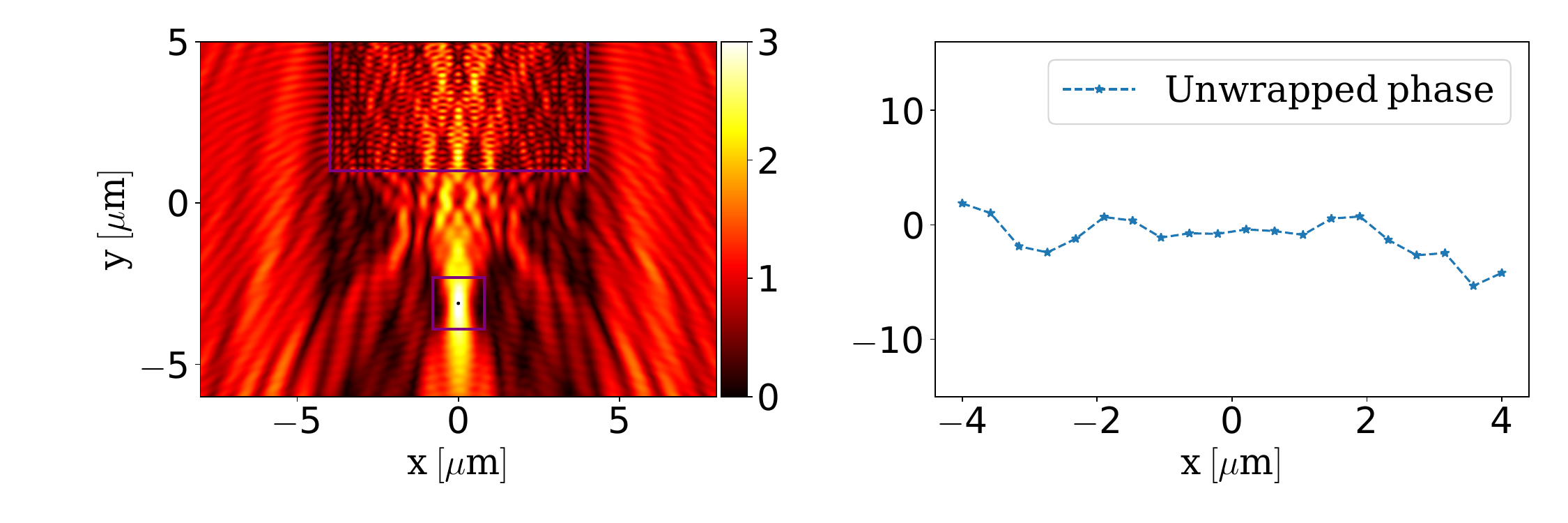}
    \caption{An optimized, 20-component phase modulation $\varphi(x)$, and the resulting $|\Etot|$. The PNJ position is 4.11 $\mu m$ along the negative $y$-axis from the micro-element boundary and at $x=0 \,\mu m$. The phase is embodied with a resolution given by $\lambda / (2 N_A)$, with $N_A$ the numerical aperture.}
    \label{fig:pos92}
\end{figure}

\begin{figure}[hbt!]
    \centering
    \includegraphics[width=0.8\textwidth]{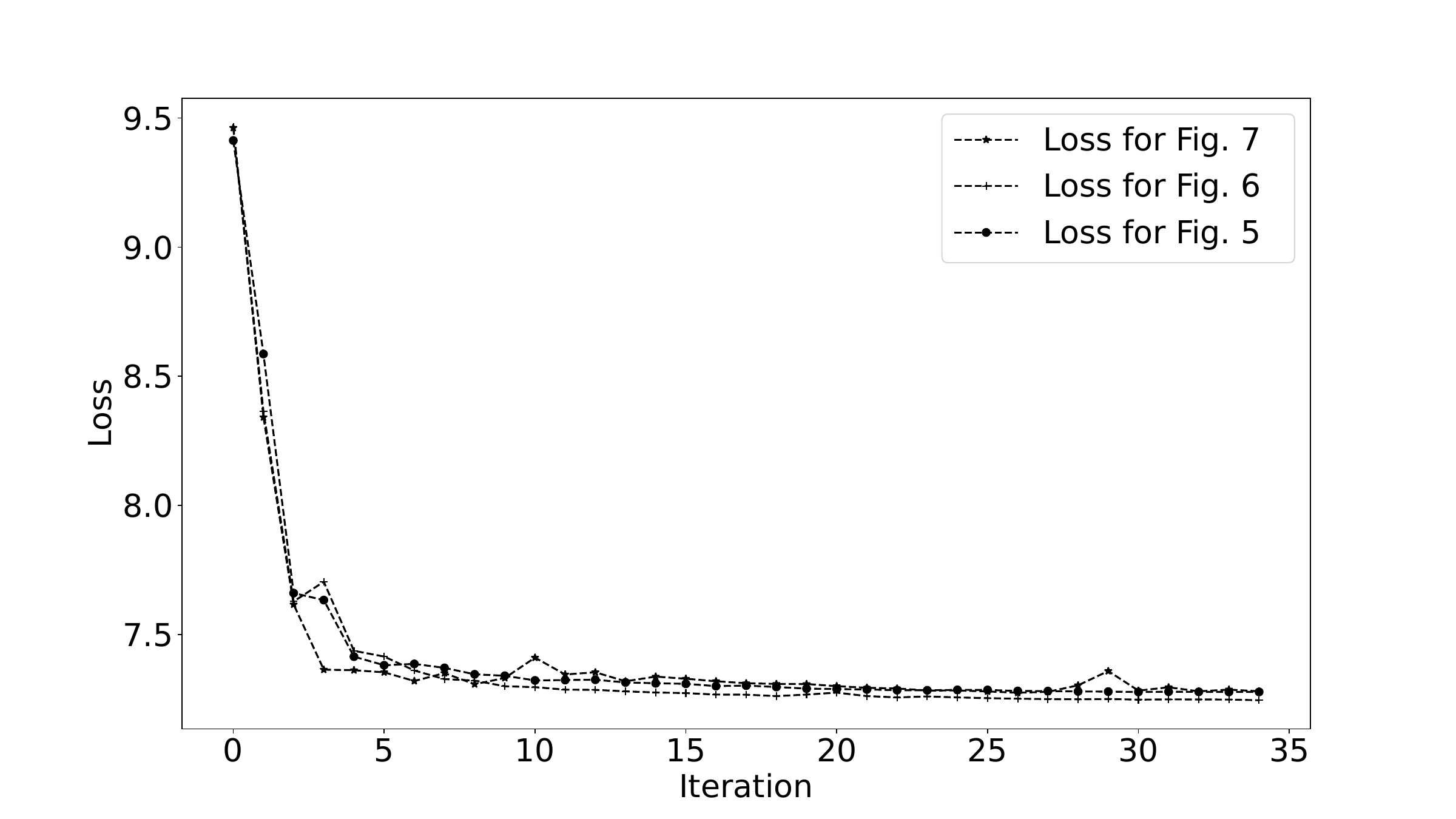}
    \caption{Optimization convergence plot for the three cases from figures \ref{fig:pos33}, \ref{fig:pos88}, and \ref{fig:pos92}.}
    \label{fig:convergence}
\end{figure}

To implement the framework, following our definition of a PNJ in Section \ref{sec:new_measure}, we used the Python Optimal Transport library \cite{flamary2021pot} for the calculation of the Wasserstein distances and relied on the package \texttt{pykan} from \cite{Liu2024} for the implementation of KANs. We used the Finite Element Method electromagnetic field solver COMSOL Multiphysics \cite{COMSOL} to compute the time-average total electromagnetic energy densities; other numerical methods relevant in scattering by microparticles are described, e.g., in~\cite{2025-Outline,2011-nearfcomp}. We investigated this framework as a case study in $\R^2$, where we chose the phase modulation $\varphi$ to depend only on the coordinate $x$ and to vary only within the micro-element range. To represent the mapping $x \mapsto \varphi(x)$ that results in the desired PNJ in a given position, we chose to use a KAN with $n_{\rm in} = 20$ input nodes, a single hidden layer of size $4$, and $n_{\rm out}=20$ nodes in the output layer. The refractive index and the side length of the microelement were $n_{\rm e}(x) = 1.406$ and $8\,\,\mu$m, respectively, and we chose the operating wavelength $\lambda_0 = 532$ nm. This lead to the results in figures \ref{fig:pos33}, \ref{fig:pos88}, and \ref{fig:pos92}. Here we see the formation of a PNJ within the chosen domains $\Omega$ in all three shown cases, together with the corresponding phase profiles of the illumination. The only explicit regularization used in our calculations is the chosen size of the domain $\Omega$. The effect of our deliberate omission of regularization is particularly apparent in Figure \ref{fig:pos92}, where the obtained field is asymmetric, and could have been symmetrized by a simple mirroring of the phase profile $\varphi$ across $x=0$. Regardless, the KAN network almost recovers the correct symmetry, and the observed field asymmetry is possibly an artifact of overfitting. For all three cases, Figure \ref{fig:convergence} indicates that the optimization process has converged.
\FloatBarrier

\section{Conclusion}\label{sec:conclusion}
We have introduced a new measure for the concentration of electromagnetic fields, leading to a new definition of the photonic nanojet. The measure and the definition are independent of a particular field shape near the point of concentration, and they are rigorously justified as well as quantifiable. As further justification, but also as a contribution of its own, we have applied our new PNJ concentration measure to achieve a machine-learned method of PNJ steering with phase-only modulated illumination and no opto-mechanical intervention.
\clearpage
\FloatBarrier

\section*{Funding}
This work was supported by research grant no. 58857 from the Villum Foundation.


\begin{thebibliography}{1}

\bibitem{Darafsheh-2021}
A. Darafsheh, \enquote{Photonic nanojets and their applications,} J. Phys. Photonics \textbf{3}, 022001 (2021).

\bibitem{Lecler-2019}
S. Lecler, S. Perrin, A. Leong-Hoi, and P. Montgomery, \enquote{Photonic Jet Lens,} Sci. Rep. \textbf{9}, 4725 (2019).

\bibitem{Itagi-2005}
A. V. Itagi and W. A. Challener, "Optics of photonic nanojets," J. Opt. Soc. Am. A \textbf{22}(12), 2847 (2005).

\bibitem{Heifetz-2006}
A. Heifetz, K. Huang, A. V. Sahakian, X. Li, A. Taflove, and B. Vadim, \enquote{Experimental confirmation of backscattering enhancement induced by a photonic jet,} Appl. Phys. Lett. \textbf{89}, 221118 (2006).

\bibitem{2023-phase-only_PNJ}
M. Karamehmedovi\'c and J. Gl\"uckstad, \enquote{Phase-only steerable photonic nanojets,} Opt. Express \textbf{31}(17), (2023).

\bibitem{2022-PNJ1}
M. Karamehmedovi\'c, K. Scheel, F. L.-S. Pedersen, A. Villegas, and P.-E. Hansen, \enquote{Steerable photonic jet for super-resolution microscopy,} Opt. Express \textbf{30}(23), 41757 (2022).

\bibitem{2022-SPIE} M. Karamehmedovi\'c, K. Scheel, F. L.-S. Pedersen,
P.-E. Hansen, \enquote{Imaging with a steerable photonic nanojet probe,} Proc.
SPIE 12203, Enhanced Spectroscopies and Nanoimaging 2022, 1220306 (3
October 2022); doi: 10.1117/12.2633442

\bibitem{Huszka-2019}
G. Huszka and M. A. M. Gijs, \enquote{Super-resolution optical imaging: A comparison,} MNE \textbf{2}(7) (2019).

\bibitem{Chen-2004}
Z. Chen and A. Taflove, \enquote{Photonic nanojet enhancement of backscattering of light by nanoparticles: a potential novel visible-light ultramicroscopy technique,} Opt. Express \textbf{12}(7), 1214 (2004).

\bibitem{Li-2005}
X. Li, Z. Chen, A. Taflove, and V. Backman, \enquote{Optical analysis of nanoparticles via enhanced
backscattering facilitated by 3-d photonic nanojets,} Opt. Express \textbf{13}(2), 526 (2005).

\bibitem{Darafsheh-2012}
A. Darafsheh, G. F. Walsh, L. Dal Negro, and V. N. Astratov, \enquote{Optical super-resolution by high-
index liquid-immersed microspheres,} Appl. Phys. Lett. \textbf{101}, 141128 (2012).

\bibitem{Rodrigo}
P. J. Rodrigo, I. R. Perch-Nielsen, C. A. Alonzo and J. Glückstad, \enquote{GPC-based optical micromanipulation in 3D real-time using a single spatial light modulator,} Opt. Express \textbf{14}(26), 13107--13112 (2006).

\bibitem{Minin-2020}
I. V. Minin, O. V. Minin, C. Y. Liu, H. D. Wei, Y. E. Geints, and A. Karabchevsky. \enquote{Experimental Demonstration of a Tunable Photonic Hook by a Partially Illuminated Dielectric Microcylinder,} Opt. Lett., \textbf{45}(17), 4899 (2020).

\bibitem{Neves-2015}
{\color{black}A. A. R. Neves, \enquote{Photonic nanojets in optical tweezers,}  J. Quant. Spectrosc. Radiat. Transf. \textbf{162}, 122 (2015).}

\bibitem{Rodrigo2}
P. J. Rodrigo, R. L. Eriksen, V. R. Daria and J. Glückstad, \enquote{Shack-Hartmann multiple-beam optical tweezers}, Opt. Express \textbf{11}(3), 208--214 (2003).

\bibitem{Sergeeva-2020}
K. A. Sergeeva, M. V. Tutov, S. S. Voznesenskiy, N. I. Shamich, A. Yu. Mironenko, and A. A.
Sergeeva, \enquote{Highly-sensitive fluorescent detection of chemical compounds via photonic nanojet
excitation,} Sens. Actuators B Chem. \textbf{305}, 127354 (2020).

\bibitem{Ruzankina-2020}
I. S. Ruzankina and G. Ferrini, \enquote{Enhancement of Raman signal by the use of BaTiO3
Microspheres,} Proceedings -- International Conference Laser Optics 2020, Iclo 2020, 9285882 (2020).
%
%
%
%
\bibitem{Sergeeva2020A}
Aleksandr Sergeev, Ksenia Sergeeva, \enquote{Functional dielectric microstructure for photonic nanojet generation in reflection mode}, Opt. Mater. \textbf{110}, 110503, ISSN 0925-3467 (2020).
%
\bibitem{Hua2023}
Yu-Jing Yang, De-Long Zhang, Ping-Rang Hua, \enquote{Array of photonic hooks generated by multi-dielectric structure}, Opt. Laser Technol. \textbf{157}, 108673 (2023).
%
\bibitem{Simon2024}
Matthies, Alex J. and Mortlock, Jonathan M. and McArd, Lewis A. and Raghuram, Adarsh P. and Innes, Andrew D. and Gregory, Philip D. and Bromley, Sarah L. and Cornish, Simon L., \enquote{Long-distance optical-conveyor-belt transport of ultracold $^{133}\mathrm{Cs}$ and $^{87}\mathrm{Rb}$ atoms}, Phys. Rev. A \textbf{109}, 023321 (2024).
%
\bibitem{Pierron2019}
Robin Pierron, Grégoire Chabrol, Stéphane Roques, Pierre Pfeiffer, Jean-Paul Yehouessi, Géraud Bouwmans, and Sylvain Lecler, \enquote{Large-mode-area optical fiber for photonic nanojet generation}, Opt. Lett. \textbf{44}, 2474--2477 (2019).
%
\bibitem{Bouaziz2021}
Bouaziz D, Chabrol G, Guessoum A, Demagh N-E, Lecler S., \enquote{Photonic Jet-Shaped Optical Fiber Tips versus Lensed Fibers}, Photonics \textbf{8}(9), 373 (2021).
%
\bibitem{Aljuaid2022}
Wasem Aljuaid, Joseph Arnold Riley, Noel Healy, and Victor Pacheco-Peña, \enquote{On-fiber high-resolution photonic nanojets via high refractive index dielectrics}, Opt. Express \textbf{30}, 43678--43690 (2022).
%
\bibitem{Vairagi2023}
Vairagi, C., Hayenga, H. N., \& Wax, A., \enquote{On-fiber Photonic Nanojet Enables Super-Resolution in en face OCT and Scattering Nanoscopy}, Commun. Phys. \textbf{8}, 89 (2023).
%
\bibitem{Yue2023}
Yue, Y., He, X., Zhu, Q., \& Fu, L., \enquote{Generation of Long Photonic Nanojet by a Self-assembled Microdevice on Optical Fiber}, Opt. Laser Technol. \textbf{159}, 109043 (2023).
%
\bibitem{Umar2023}
Umar, M., Ozek, E. A., Abdul, B., Shafaghi, A. H., \& Yapici, M. K.,  \enquote{Extremely long nanojet formation from a ballpoint photonic pen}, JOSA B \textbf{40}(2), 284--292 (2023).
%
\bibitem{Li2024}
Li, L. P.-H., Hung, T.-Y., Chen, W.-Y., Chung, H.-J., Cheng, C.-H., Chang, T.-L., Chen, Y.-B., Minin, O. V., Minin, I. V., \& Liu, C.-Y.,  \enquote{Direct laser micro-drilling of high-quality photonic nanojet achieved by optical fiber probe with microcone-shaped tip}, Appl. Phys. A \textbf{131}, 16 (2024).

%
%
%
\bibitem{Liu-2021}
C. Y. Liu, W. Y. Chen, Y. E. Geints, O. V. Minin, and I. V. Minin, \enquote{Simulation and Experimental Observations of Axial Position Control of a Photonic Nanojet by a Dielectric Cube with a Metal Screen}, Opt. Lett. \textbf{46}(17), 4292 (2021).
%
%
%
\bibitem{Sergeeva2020B}
Sergeeva, E. A., Zelenina, A. S., Arkhipova, V. N., \& Zelenin, M. M., \enquote{Photonic nanojet-enhanced fluorescence detection of biomolecular analytes}, Sens. Actuators B: Chem. \textbf{320}, 128361 (2020).
%
\bibitem{LocalizedPNJ2022}
Zhang, Y., Liu, H., Wang, J., \& Zhang, L., \enquote{Localized photonic nanojet sensing platform based on semi-open microwell structures}, Biosens. Bioelectron. \textbf{208}, 114224 (2022).
%
\bibitem{Xiong2024}
Xiong, J., Huang, Y., Yang, X., Li, R., \& Sun, M., \enquote{Photonic nanojet-assisted selective photoactivation in live cells using a microlens-tipped fiber}, PhotoniX \textbf{5}, Article 42 (2024).
%
%
%
\bibitem{GluckstadMadsen-2023}
J. Gl\"uckstad and A. E. G. Madsen, \enquote{New analytical diffraction expressions for the Fresnel–Fraunhofer transition regime,} Optik \textbf{285}, 170950 (2023).

\bibitem{2018-bandwidth}
M. Karamehmedovi\'c, \enquote{Explicit tight bounds on the stably recoverable information for the inverse source problem,} J. Phys. Commun. \textbf{2}, 095021 (2018).

\bibitem{Paganini-2015}
A. Paganini, S. Sargheini, R. Hiptmair, and C. Hafner, \enquote{Shape optimization of microlenses,} Opt. Express \textbf{23}(10), 13099 (2015).

\bibitem{2024-localization_improved_concept}
M. Karamehmedovi\'c and F. Triki, \enquote{Localization Coefficients of Functions with
Applications in Partial Differential Equations,} \textit{submitted}, arXiv:2504.12033 

\bibitem{Balanis}
C. A. Balanis, \textit{Advanced Engineering Electromagnetics,} 2nd Ed., Wiley 2012.

\bibitem{Villani}
C. Villani, Optimal Transport: Old and New, vol. 338, Springer, 2009.

\bibitem{flamary2021pot}
Rémi Flamary, Nicolas Courty, Alexandre Gramfort, Mokhtar Z. Alaya, Aurélie Boisbunon, Stanislas Chambon, Laetitia Chapel, Adrien Corenflos, Kilian Fatras, Nemo Fournier, Léo Gautheron, Nathalie T.H. Gayraud, Hicham Janati, Alain Rakotomamonjy, Ievgen Redko, Antoine Rolet, Antony Schutz, Vivien Seguy, Danica J. Sutherland, Romain Tavenard, Alexander Tong, Titouan Vayer,
POT Python Optimal Transport library,
JMLR (2021).

\bibitem{Liu2024}
Ziming Liu and Yixuan Wang and Sachin Vaidya and Fabian Ruehle and James Halverson and Marin Soljačić and Thomas Y. Hou and Max Tegmark, \enquote{KAN: Kolmogorov-Arnold Networks,} arXiv:2404.19756.

\bibitem{COMSOL}
COMSOL Multiphysics\textregistered{ }v. 6.2. www.comsol.com. COMSOL AB, Stockholm, Sweden.

\bibitem{2025-Outline}
H. Hooshmand, T. Pahl, A. Birk, L. Fu, P.-E. Hansen, M. Karamehmedović, R. Leach, P. Lehmann, and S. Piano, \enquote{Comparison of established rigorous scattering models to accurately replicate the behavior of scattered electromagnetic wave,} J. Comput. Phys. \textbf{521}, 113519 (2025).

\bibitem{2011-nearfcomp}
M. Karamehmedovi\'c, R. Schuh, V. Schmidt, T. Wriedt, C. Matyssek, W. Hergert, A. Stalmashonak, G. Seifert, and O. Stranik, \enquote{Comparison of numerical methods in near-field 
computation for metallic nanoparticles,} Opt. Express \textbf{19}(9), (2011).

\end{thebibliography}
\end{document}